\documentclass[pra,twocolumn,preprintnumbers,amsmath,amssymb,showpacs,nofootinbib,floatfix]{revtex4}

\usepackage{graphicx,bm}

\newcommand{\nuxy}{{\nu_{\textsc{xy}}}}
\newcommand{\oxy}{{\omega_{\textsc{xy}}}}

\makeatletter
\def\graphicscale{\twocolumn@sw{0.3}{0.4}}
\def\graphicthreescale{\twocolumn@sw{0.3}{0.4}}

\begin{document}

\title{Phase diagram and multicritical behaviors of mixtures of 3D
  bosonic gases}

\author{Giacomo Ceccarelli,$^1$ Jacopo Nespolo,$^{1,2}$ 
Andrea Pelissetto,$^3$ and Ettore Vicari$^1$} 

\address{$^1$ Dipartimento di Fisica dell'Universit\`a di Pisa and
  INFN, Largo Pontecorvo 3, I-56127 Pisa, Italy} 
\address{$^2$
Department of Physics, Arnold Sommerfeld Center for Theoretical Physics, and 
Center for Nanoscience, Ludwig-Maximilians-Universit\"at M\"unchen, 
Theresienstrasse 37, D-80333 Munich, Germany}
\address{$^3$
  Dipartimento di Fisica dell'Universit\`a di Roma ``La Sapienza'' and
  INFN, Sezione di Roma I, I-00185 Roma, Italy}

\date{\today}

\begin{abstract}

We investigate the Bose-Einstein condensation patterns, the critical
and multicritical behaviors of three-dimensional mixtures of bosonic
gases with short-range density-density interactions. These systems
have a global U(1)$\oplus$U(1) symmetry, as the system Hamiltonian is
invariant under independent U(1) transformations acting on each
species.  In particular, we consider the three-dimensional
Bose-Hubbard model for two lattice bosonic gases coupled by an on-site
inter-species density-density interaction.  We study the phase diagram
and the critical behaviors along the transition lines characterized by
the Bose-Einstein condensation of one or both species. We present
mean-field calculations and numerical finite-size scaling analyses of
quantum Monte Carlo data. We also consider multicritical points, close
to which it is possible to observe the condensation of both gas
components.  We determine the possible multicritical behaviors by
using field-theoretical perturbative methods. We consider the
U(1)$\oplus$U(1)-symmetric Landau-Ginzburg-Wilson $\Phi^4$ theory and
determine the corresponding stable fixed points of the
renormalization-group flow. The analysis predicts that, in all cases,
the multicritical behavior is analogous to the one that would be
observed in systems of two identical gases, with an additional
$\mathbb{Z}_2$ exchange symmetry.
\end{abstract}

\pacs{67.25.dj,67.85.Hj,05.70.Jk,05.10.Cc}

\maketitle



\section{Introduction}
\label{intro}

The complex behavior of mixtures of bosonic gases 
has been extensively investigated experimentally---in
particular, in cold-atom systems~\cite{PhysRevLett.78.586,
  PhysRevLett.81.1539,Nature.396.345,
  PhysRevLett.82.2228,PhysRevLett.85.2413,PhysRevA.63.051602,
  PhysRevLett.89.053202,PhysRevLett.89.190404,PhysRevLett.99.190402,
  PhysRevLett.99.010403,PhysRevA.77.011603,PhysRevLett.100.210402,
  PhysRevLett.101.040402,PhysRevA.79.021601,
  PhysRevLett.103.245301,PhysRevA.80.023603,PhysRevLett.105.045303,
  PhysRevA.82.033609,PhysRevA.84.011610,NatPhys7.434,NatPhys8.71} ---and
theoretically~\cite{HS-96,Boninsegni-01,AHDL-03,DDL-03,KS-03,PC-03,
  KG-04,KPS-04,ICSG-05,PSP-08,SCPS-09,HSH-09,CSPS-10,
  FHRSB-11,FFPP-11,Pollet-12,PFFP-12,ACV-14,LCD-14,GJABBP-14,
  GBS-15,SMSR-15,CNPV-15,LCGP-15}. These systems exhibit a rich
behavior, at zero and finite temperature, with several different 
phases separated by transition lines, along which one or more components
of the system undergo Bose-Einstein condensation (BEC).

In this paper we consider three-dimensional (3D) mixtures of two
different bosonic gases with short-range interactions that only depend
on the local densities of the two gases. The Hamiltonian of these
systems is invariant under U(1) transformations acting independently
on each species, so that the model is U(1)$\oplus$U(1) symmetric.  In
particular, we consider the 3D two-component Bose-Hubbard model with
an on-site inter-species density-density interaction.  This is a
realistic model for two bosonic species in optical
lattices~\cite{BDZ-08}.

We determine the finite-temperature phase diagram by using a variety
of techniques. First, we consider the mean-field (MF) approximation,
determining the qualitative phase behavior of the system as a function
of the model parameters, such as the chemical potentials and the
on-site inter- and intra-species couplings.  We find several different
phases, in which each species may be in the normal or superfluid
state, and identify critical lines and multicritical points (MCPs)
where some transition lines meet.

The 3D phase diagram is investigated in the hard-core (HC) limit of
each species by a finite-size scaling (FSS) analysis of quantum Monte
Carlo (QMC) simulations.  The numerical data allows us to identify the
universality class of the transition lines that correspond to the BEC
of one of the two species.  We show that, independently whether the
noncritical component is in the normal or superfluid phase, the
critical behavior of the condensing species belongs to the 3D XY
universality class characterized by the breaking of a global U(1)
symmetry and by short-range effective interactions.  This is the same
universality class associated with the BEC of a single bosonic
gas~\cite{CPS-07,CR-12,CTV-13,CN-14} (and also with the superfluid
transition in $^4$He~\cite{Lipa-etal-96,CHPV-06}, with transitions in
some liquid crystals characterized by density or spin waves and in
magnetic systems with easy-plane anisotropy, etc.~\cite{PV-02}).  This
result implies an effective decoupling of the critical modes of the
condensing species from those of the noncritical component,
independently whether the latter is in the normal or superfluid phase.

The phase diagram of mixtures of bosonic gases also presents
particular points where some transition lines meet.  Multicritical
behaviors develop at these MCPs, arising from the competition of the
two U(1) order parameters associated with the BEC of the two species.
To identify the possible universality classes of the multicritical
behaviors, we use the field-theoretical approach, considering the
effective Landau-Ginzburg-Wilson (LGW) $\Phi^4$ theory for two complex
fields with global U(1)$\oplus$U(1) symmetry.  We study the
renormalization-group (RG) flow in the quartic-parameter space,
identifying the stable fixed points (FPs), which control the critical
behavior, and their attraction domain.

The paper is organized as follows.  In Sec.~\ref{model} we define the
Bose-Hubbard model for two lattice bosonic gases.  In Sec.~\ref{phmf}
we determine the phase diagram of the model in the MF approximation,
showing that, by changing the model parameters, one can obtain
qualitatively different behaviors.  In Sec.~\ref{critbeh} we present
our numerical results and determine numerically the critical behavior
along the transition lines where one species undergoes BEC. In
Sec.~\ref{multi} we study the multicritical behaviors at MCPs where
some transition lines meet in the phase diagram.  Finally, in
Sec.~\ref{conclu} we draw our conclusions. App.~\ref{5loop} reports
the five-loop perturbative series of the minimal-subtraction scheme,
which are used in the RG study of the multicritical behavior.

\section{The Bose-Hubbard model of a mixture of bosonic gases}
\label{model}

The two-species Bose-Hubbard (2BH) model is a lattice model 
appropriate to describe mixtures of bosonic gases with local 
density-density  interactions. Its Hamiltonian reads
\begin{eqnarray}
H &=& - \sum_{s,\langle {\bm x \bm y} \rangle}
t_s (b_{s\bm{x}}^\dagger b_{s\bm{y}}^{} + {\rm h.c}) - \sum_{s,\bm{x}}
\mu_s \, n_{s\bm{x}}
\label{sBH}\\
&+& \frac{1}{2} 
 \sum_{s,\bm{x}} V_s \,n_{s\bm{x}} (n_{s\bm{x}}-1)  + 
U \sum_{\bm{x}} n_{1\bm{x}} n_{2\bm{x}},
\nonumber
\end{eqnarray}
where ${\bm x}$ is a site of a cubic lattice, $\langle {\bm x \bm y}
\rangle$ labels a lattice link connecting two nearest-neighbor sites,
the subscript $s$ labels the two species, and $n_{s\bm{x}}\equiv
b_{s\bm{x}}^\dagger b_{s\bm{x}}^{}$ is the density operator of the
$s$-species.  The 2BH model is symmetric under the U(1)
transformations $b_{s\bm{x}} \to e^{i\theta_s} b_{s\bm{x}}$ acting
independently on the two species.  Therefore, the global symmetry
group is ${\rm U(1)} \oplus {\rm U(1)}$.

For $t_1=t_2$, $\mu_1=\mu_2$, and $V_1=V_2=V$, the 2BH model
(\ref{sBH}) describes the behavior of a mixture of two identical
bosonic gases and it has been extensively studied in
Ref.~\cite{CNPV-15}.  In this case, the model has an additional
${\mathbb Z}_2$ exchange symmetry, so that the symmetry group becomes
${\mathbb Z}_{2,e} \otimes [{\rm U}(1) \oplus {\rm U}(1)]$.  In the HC
limit, or, more generally, when $V \gtrsim U$, both components
condense at the transition.  Thus, the global symmetry breaks to
${\mathbb Z}_{2,e} \otimes [{\mathbb Z}_2 \oplus {\mathbb Z}_2]$. The
critical behavior is controlled by a decoupled 3D XY fixed point
\cite{CNPV-15}: The critical behaviors of the two gases are
effectively decoupled and belong to the 3D XY universality class
associated with the symmetry breaking ${\rm U(1)}\to {\mathbb
  Z}_2$. Although the inter-species density-density interaction does
not change the leading critical behavior, it plays an important role
close to criticality, as it gives rise to very-slowly-decaying scaling
corrections, which are not present at the BEC transition of a single
bosonic species.  In the opposite case, i.e., for $V \lesssim U$, only
one component condenses, so that the global symmetry is broken to
${\rm U}(1) \oplus {\mathbb Z}_2$.  The associated critical behavior
belongs to a different 3D universality class \cite{CNPV-15}.

In the following we consider mixtures of nonidentical gases described
by the 2BH model (\ref{sBH}).  As we shall see, their
finite-temperature phase diagrams present several phases where the two
species are in the normal or superfluid state, separated by transition
lines along which only one species condenses. Moreover, we will
discuss MCPs, which are points in the phase diagram where some
transition lines meet.

The hard-core (HC) limit for the $s$ component is obtained by taking
$V_s\to\infty$. In this limit, using the particle-hole transformation,
we can relate the spectrum of the Hamiltonian for two different sets
of parameters. For instance, assume that $V_1 = \infty$, so that
$n_{1\bm{x}}$ can only assume the values 0 and 1. Under a
particle-hole transformation the kinetic term is unchanged, while
$n_{1\bm{x}}\to 1 - n_{1\bm{x}}$. Thus, the spectrum of the model with
chemical potentials $\mu_1$ and $\mu_2$ and interaction $U$ is related
to that of the model with chemical potentials $\mu_1'$ and $\mu_2'$
and interaction $U'$ with
\begin{equation}
\mu_1' = -\mu_1,\quad
\mu_2' =  \mu_2 - U,\quad U' = -U.
\label{transf}
\end{equation}
Indeed, the energy levels of the two models differ by an irrelevant
constant term proportional to $\mu_1$. An analogous relation holds for
the second species if $V_2 = \infty$. These relations imply that, if
one of the two components has a hard core, one can limit oneself to
study the phase diagram for $U>0$.

\section{Mean-field phase diagrams}
\label{phmf}

\begin{figure}
\centering \includegraphics{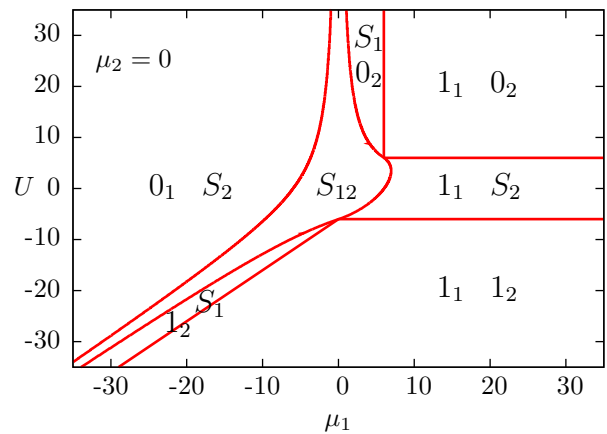}
\caption{MF zero-temperature phase diagram of the 2BH
  model in the HC limit for $\mu_2 = 0$, as a function of $\mu_1$ and 
  and $U$.  The symbol $S$ marks the
  superfluid phases: $S_1$ and $S_2$ indicate the phases in which only 
  components 1 and 2 are superfluid, respectively; in phase $S_{12}$ both 
  components are superfluid. The
  symbols $0_s$ and $1_s$ mark Mott phases for species $s$ with
  filling 0 and 1, respectively.  In the leftmost phase marked 
  $0_1, S_2$, we have $\rho_2 = 1/4$ for the superfluid density 
  and $n_2 = 1/2$, which are the
  values corresponding to a single gas with vanishing chemical
  potential.}
    \label{fig:T0_hc}
\end{figure}

\begin{figure}
    \includegraphics{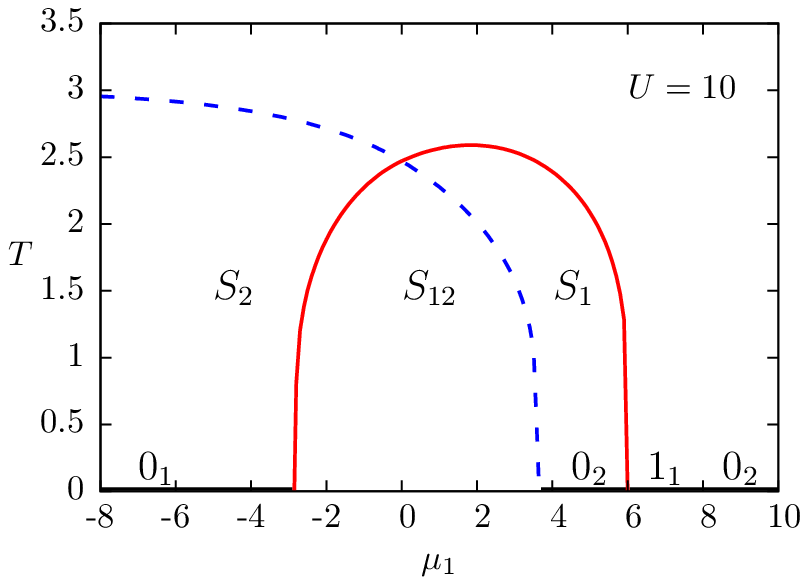}
    \includegraphics{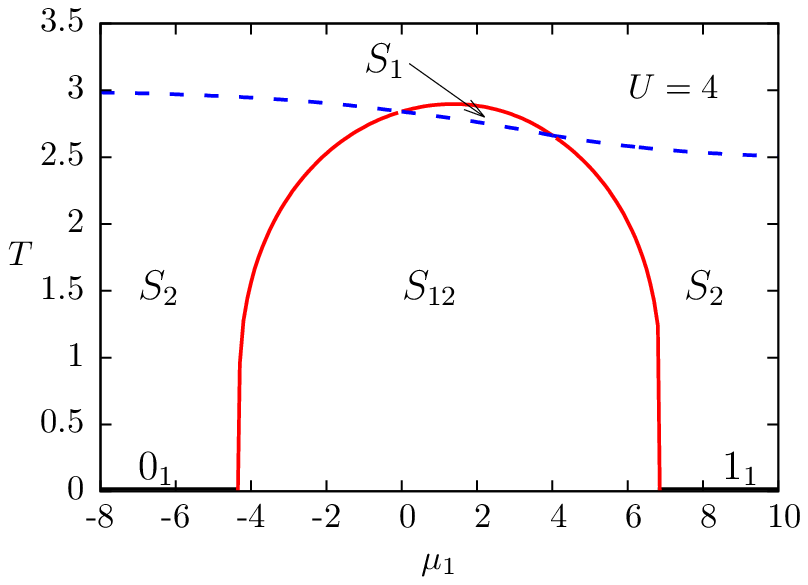}
\caption{Phase diagram of the HC 2BH model
  for $\mu_2=0$ and two values of $U$, $U=10$ (top) and $U=4$
  (bottom), as a function of $\mu_1$ and $T$.  
  In phase $S_1$ component 1 is in the superfluid state,
  while component 2 is in the normal state. In phase $S_2$ the opposite occurs.
  In phase $S_{12}$ both gases are superfluid. In the upper (and, for 
  $U=10$, rightmost) region with no
  label both gases are in the normal state. The full and dashed lines
  indicate the normal-to-superfluid transition lines of gas 1 and
  2, respectively.  }
\label{fig:Tpos_hc}
\end{figure}

Some qualitative or semi-quantitative aspects of the phase diagram can
be inferred by MF calculations. For this purpose we make the approximation
\begin{eqnarray}
  b^\dag_{s\bm{x}} b_{s\bm{y}}
  &=& \left[ (b^\dag_{s\bm{x}} - \phi_s^*) + \phi_s^* \right] 
     \left[ (b_{s\bm{y}} - \phi_s) + \phi_s \right] \nonumber \\ 
  &\approx& \phi_s b^\dag_{s\bm{x}} + \phi_s^* b_{s{\bm y}} - |\phi_s|^2,
\label{approx-meanfield}
\end{eqnarray}
where $\phi_{s}=\langle b_{s \bm x }\rangle$ are two complex
space-independent variables, which play the role of order parameters
at the BEC transitions.  The parameters $\phi_s$ are related to the
superfluid densities by $\rho_s \propto |\phi_s|^2$.
Eq.~(\ref{approx-meanfield}) allows us to rewrite Hamiltonian
(\ref{sBH}) as a sum of decoupled one-site Hamiltonians
\begin{eqnarray} 
 H_{\rm mf} &=& - 2d \sum_{s} t_s \left(\phi_s b^\dag_{s} + \phi_s^*
  b^{}_{s} - |\phi_s|^2 \right) \label{eq:mf_hamiltonian} \\ &-&
  \sum_{s} \mu_s n_{s} + \frac{1}{2} \sum_s V_s n_{s} (n_s - 1) + U
  n_{1}n_{2},  \nonumber
\end{eqnarray} 
where $n_s=b^\dag_s b_s$.  Since the spectrum of the theory is
invariant under $b_s \to e^{i\theta_s} b_s$, where $\theta_s$ are two
independent phases, the two parameters $\phi_s$ can be assumed to be real without loss of
generality.  They are determined by minimizing the single-site free
energy
\begin{equation}
\label{eq:mf_free_energy}
    F = - T  \ln \sum_i e^{-\beta E_i} \, ,
\end{equation}
where $E_i$ are the eigenvalues of $H_{\rm
  mf}$~\cite{footnote-meanfield}. At zero temperature the minimization
of the free energy corresponds to the minimization of the ground-state
energy $E_0$.  

In the following, we restrict
ourselves to the case $t_1=t_2=t$ and $V_1 = V_2 = V$. Moreover,
we set $t=1$, so that all energies are expressed in units of $t$.

The model shows a complex phase diagram, with transition lines
(surfaces) along which one component undergoes a transition from the
normal state to a superfluid one. Note that, in the limit of zero
temperature, the normal phases become Mott insulating phases or simply
correspond to the vacuum.  In the following we present MF results for
some selected values of the model parameters, which should be
representative of the different finite-temperature behaviors that can
be observed by varying the parameters.

To begin with, we consider the HC limit $V\to \infty$.
Fig.~\ref{fig:T0_hc} shows the zero-temperature phase diagram as a
function of $\mu_1$ and $U$, when component 2 has zero chemical
potential, i.e., for $\mu_2=0$. We observe several Mott and superfluid
phases, separated by continuous transition lines.  The Mott phases
appearing in Fig.~\ref{fig:T0_hc} are somewhat trivial, as they
correspond to the vacuum or to unit filling (one particle per site).
If one chooses more general values for $t_s$ and $V_s$, one may obtain
more complex Mott phases at $T=0$. For instance, for $\mu_1=\mu_2$ and
$t_1\neq t_2$, the phase diagram also shows a Mott phase with global
unit filling ($n_1+n_2=1$ but $n_s\neq 0,\,1$) and, therefore a
degenerate ground state \cite{AHDL-03}.  The large degeneracy of the
ground state may be described in terms of {\em isospin} degrees of
freedom per site, interacting by means of an effective low-energy spin
Hamiltonian~\cite{KS-03,DDL-03,AHDL-03,ICSG-05}.

At finite temperature the vacuum and Mott phases are replaced by
normal phases.  As suggested by the $T=0$ phase diagram of
Fig.~\ref{fig:T0_hc}, we may have different behaviors depending on the
strength of the inter-species interaction $U$. For example,
Fig.~\ref{fig:Tpos_hc} shows the phase diagram for $\mu_2=0$ as a
function of $\mu_1$ and $T$, for two values of $U$, $U=10$ and
$U=4$. For $U=10$ there is a single multicritical point (MCP) where
four transition lines meet. For $U=4$, instead, two different MCPs are
present. The change of behavior occurs for $U=6$. It can be related to
the phase boundary $U=2d$ appearing in the phase diagram of a single
HC bosonic gas.  In the MF approximation all phase boundaries
correspond to continuous transitions.

\begin{figure}
    \includegraphics{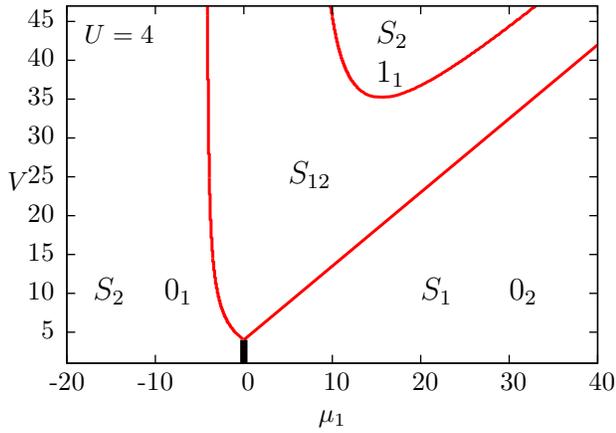}
\caption{Zero-temperature phase diagram for $\mu_2=0$ and $U=4$, as 
  a function of $\mu_1$ and $V$.
  The different phases are labelled as in
  Fig.~\ref{fig:T0_hc}.  For $V<U=4$, a first-order transition line
  (indicated by a thick line) runs along $\mu_1 = 0$.
   The phase $1_1$, $S_2$ occurs only for $V\gtrsim 35.3$.  
}
    \label{fig:T03_sc}
\end{figure}

\begin{figure}[tb]
    \includegraphics{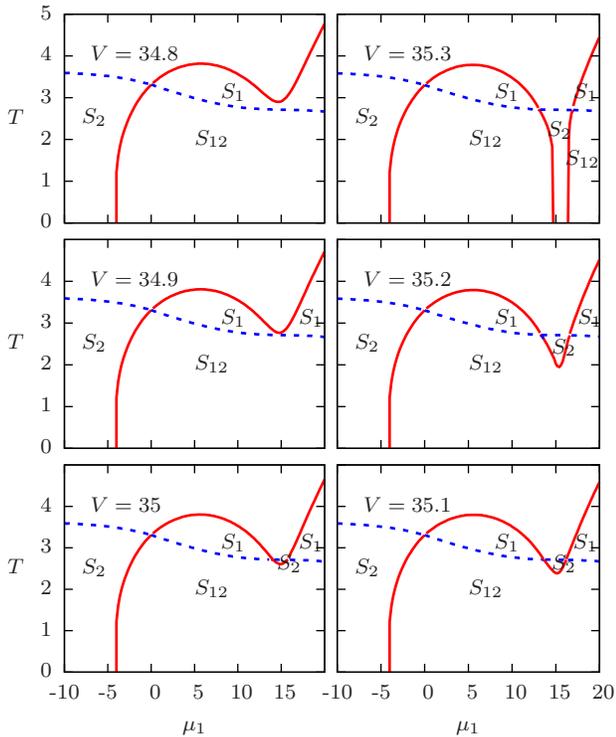}
\caption{Phase diagrams for $\mu_2=0$ and $U=4$ as a function of 
  $T$ and $\mu_1$. We report results for several 
  values of the intra-species $V$ repulsion, for $V$ close to
  $V\approx 35$ (at $T=0$, the phase $1_1$, $S_2$ occurs only for 
  $V\gtrsim 35.3$, see Fig.~\ref{fig:T03_sc}).  
  The different phases and the transition lines are labelled as in 
  Fig.~\ref{fig:Tpos_hc}. 
  }
\label{fig:Tpos1_sc}
\end{figure}

In order to understand the role of finite intra-species couplings, we
consider a finite $V$. In this case, Mott phases with higher integer
fillings are possible.  Moreover, when $V<U$ one may have first-order
transition lines between the superfluid phases of the two components.
For example, Fig.~\ref{fig:T03_sc} shows the zero-temperature behavior
for $\mu_2=0$ and $U=4$ as a function of $V$ and $\mu_1$. In this
case, the phase $S_1$, in which component 1 is superfluid and
component 2 is depleted, is separated from phase $S_2$, in which
component 2 is superfluid and component 1 is depleted, by a
first-order transition line with $\mu_1 = 0$ and $0< V < U$.

Finite intra-species interactions lead to significant changes of the
phase diagram also at finite temperature.  For example,
Fig.~\ref{fig:Tpos1_sc} shows the phase behavior for $U=4$, $\mu_2 =
0$, and some finite values of $V$, as a function of $\mu_1$ and $T$.
It should be compared with Fig.~\ref{fig:Tpos_hc}, where we report the
phase diagram in the HC limit for the same values of $U$ and $\mu_2$.
As $V$ is increased from $V=34.8$ to $V=35.3$, the phase diagram
changes qualitatively.  For $V\lesssim 35$ only one MCP is present,
while for $V\gtrsim 35$ three MCPs occur. Moreover, for $V\approx
35.3$ there are two different $S_{12}$ phases. In particular, both
components condense for any $\mu_1\gtrsim 17$, if $T$ is not too
large.  Such a large-$\mu_1$ $S_{12}$ phase occurs for any finite $V$:
for $\mu_1$ larger than a $V$ dependent value $\mu_b(V)$, i.e., for
$\mu_1\gtrsim \mu_b(V)$, both components always condense for small
$T$.  For $V\to\infty$ we have $\mu_b(V) \to \infty$, so that such
phase does not exist in the HC case, as shown in the lower panel of
Fig.~\ref{fig:Tpos_hc}.

\begin{figure}
    \includegraphics{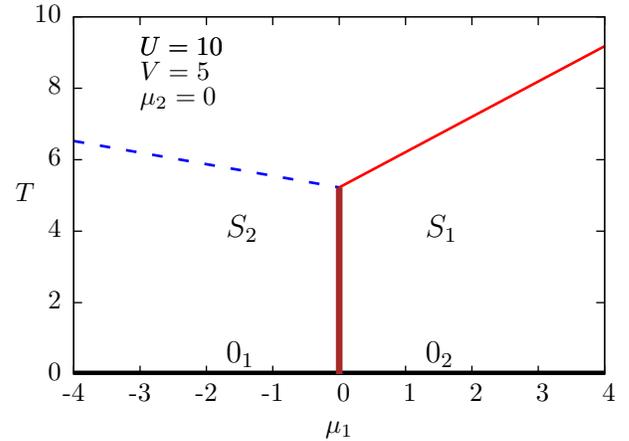}
\caption{Phase diagram for $\mu_2=0$, $U=10$, and $V=5$, as function
  of $\mu_1$ and $T$.  Species 1 and 2 are superfluid for $\mu_1>0$
  and $\mu_1<0$, respectively. These phases are separated by a
  first-order transition line along $\mu_1=0$, ending at the point
  where the continuous normal-to-superfluid transition lines of the
  two gases meet.  In the uppermost phase with no labels, both
  components are in the normal state.  }
    \label{fig:fst_order}
\end{figure}

When $V<U$ the finite-temperature phase diagram changes significantly from
that observed in the regime $V>U$. Indeed, as shown in
Fig.~\ref{fig:fst_order}, in the phase diagram for $\mu_2=0$, $V=5$,
and $U=10$ three transition lines meet at a MCP: two continuous
normal-to-superfluid transition lines and a first-order transition
line separating the superfluid phases of gases 1 and 2 along the
line $\mu_1=0$. This is of course consistent with what observed 
at zero temperature, see Fig.~\ref{fig:T03_sc}.

MF calculations can be straightforwardly extended to more general
cases, such as $V_1\neq V_2$ and/or $t_1\neq t_2$. However, 
the main features of the possible finite-temperature behaviors
should be already present in the results shown above.

\section{Numerical results}
\label{critbeh}

In this section we investigate numerically the nature of the
transitions occuring in two-component bosonic systems described by the
Hamiltonian (\ref{sBH}).  As already discussed in the MF
approximation, in most of the cases the transition lines are
associated with normal-to-superfluid transitions of one of the two
species. However, it is also possible to have first-order transition
lines between two phases in which only one component is superfluid
(this behavior is expected in the soft-core regime, see
Fig.~\ref{fig:fst_order}) and MCPs where three or four transition
lines meet.

Since the normal-to-superfluid transition of a single species is
related to the spontaneous breaking of the U(1) symmetry of the
condensing species, it seems natural that the critical behavior
belongs to the standard 3D XY universality class~\cite{PV-02}.  This
conjecture, however, requires that the second (spectator) component
plays no role at the transition, i.e., an effective decoupling of the
critical modes of the condensing species from the modes of the
spectator one.  This hypothesis is quite natural when the spectator
species is in the normal phase, which is characterized by short-range
correlations. However, if the second component is in the superfluid
phase, in which long-range spin-wave (Goldstone) modes develop, the
asymptotic decoupling of the two species is no longer obvious. Indeed,
the Goldstone modes may give rise to effective long-range interactions
among the condensing particles.  As a consequence, one might observe a
different critical behavior (we recall that the standard 3D XY
universality is observed only if the interactions decay sufficiently
fast with the distance), or XY behavior with peculiar slowly-decaying
scaling corrections, as it occurs in the case of mixtures of identical
bosonic gases~\cite{CNPV-15}.

To investigate these issues, we perform quantum Monte Carlo (QMC)
simulations of the 2BH model in the HC limit for both species
($V_1=V_2\to\infty$).  Our results provide a robust evidence that the
critical behaviors belong to the 3D XY universality class along all
continuous transition lines, including those where one species
condenses in the superfluid background of the other one. In other
words, the local inter-species density-density interaction is an
irrelevant RG perturbation at all BEC transition lines. There is also
no evidence of slowly-decaying scaling corrections.  Apparently, the
leading scaling corrections are always controlled by the leading
irrelevant RG operator which appears in the standard XY model or at
the BEC transition of a one-component bosonic gas.

As we shall see in Sec.~\ref{multi}, these features change when we
approach a MCP, where several transition lines meet. In that case the
competition of the two condensing order parameters gives rise to more
important effects.

\subsection{QMC simulations}
\label{qmcsi}

We consider the 2BH model (\ref{sBH}) with $t_1=t_2=t=1$ in the HC
limit $V_1,V_2\to\infty$, for cubic $L^3$ lattices with periodic
boundary conditions. We perform QMC simulations~\cite{SK-91,SS-02} for
$\mu_2=0$ and two values of $U$, $U=10$ and $U=4$, using the same
algorithm employed in Ref.~\cite{CNPV-15} (we refer to this reference
for technical details).  In the MF approximation, the phase behavior
as a function of $T$ and $\mu_1$ is reported in
Fig.~\ref{fig:Tpos_hc}. Here, we verify that the MF diagram is
qualititavely correct. Moreover, we determine the nature of the
critical behavior at a few selected points, at which the spectator
species is both in the normal and in the superfluid phase.

For this purpose, we focus on the finite-size scaling (FSS) 
behavior of the helicity modulus,
which generally provides the most precise numerical results to
characterize the critical behavior.  The helicity
modulus $\Upsilon_s$ of species $s=1,2$ is obtained from the response
of species $s$ to a twist in the boundary conditions by an angle
$\alpha_s$, i.e.,
\begin{equation}\label{eq:helicity_def}
  \Upsilon_s \equiv - \frac{1}{L} \left.
  \frac{\partial^2 Z(\alpha_s)}{\partial\alpha_s^2}
  \right|_{\alpha_s = 0} , 
\end{equation}
where $Z(\alpha_s)$ is the partition function for twisted boundary 
conditions in one direction and periodic boundary conditions in the 
two orthogonal directions.  In QMC simulations $\Upsilon_s$
is simply related to the linear winding number $W_s$ of species $s$,
through the relation $\Upsilon_s = \langle W_s^2 \rangle/L$.

We also computed expectation values of other observables, such as the
two-point function $\langle b^{\dag}_{s{\bf x}} b_{s{\bf y}} \rangle$,
its spatial integral, and the second-moment correlation length. In the
following we do not report the corresponding results. 
We only mention that they substantially confirm the conclusions 
drawn from the analysis of the helicity modulus.

The helicity modulus at the BEC transition of the $s$-species is
expected to behave as
\begin{equation}
  R_s \equiv \Upsilon_s L \approx f(u L^{1/\nu}),
\label{eq:Rscaling}
\end{equation}
where $\nu$ is the correlation-length exponent, and the linear scaling
field $u$ is a linear combination of $T$ and of the model parameters,
which vanishes at the critical point.  Assuming for simplicity that
$\mu_2$ is fixed, at a generic critical point $(T_c,\mu_{1c})$, the
scaling field can be written as
\begin{equation}
u(T,\mu_1) \approx  a (T-T_c) + b (\mu_1 - \mu_{1c}),
\label{usca}
\end{equation}
where $a$ and $b$ are nonuniversal coefficients.  Thus, if we fix
$\mu_1$ to its critical value, that is we set $\mu_1 = \mu_{1c}$, and
we investigate the transition by varying $T$, we have simply
$u=a(T-T_c)$. If instead $T$ is fixed to its critical value, we have
$u=b(\mu_1-\mu_{1c})$.  The scaling function $f(x)$ is universal,
provided that coefficients $a$ and $b$ appearing in Eq.~(\ref{usca})
are properly defined.  However, it depends on the shape and boundary
conditions of the system.  A straightforward consequence of
Eq.~(\ref{eq:Rscaling}) is that the curves $R_s(L;\mu_1)$ at fixed $L$
cross each other at the critical point, where their slopes are
controlled by the correlation-length exponent $\nu$.

We wish now to verify our conjecture that the critical behavior of the
condensing component always belongs to the 3D XY universality class,
irrespective of the (normal or superfluid) state of the spectator
component. If true, the exponent $\nu$ appearing in
Eq.~(\ref{eq:Rscaling}) equals that of the 3D XY universality
class~\cite{CHPV-06} $\nuxy=0.6717(1)$.  Moreover, also the scaling
function $f(x)$ must be equal to that of the 3D XY universality class
apart from a trivial multiplicative rescaling of the argument.  In
particular, we should find~\cite{CHPV-06} $R^*=f(0) =0.516(1)$ at the
transition.

We should note that an accurate determination of the critical
parameters requires us to take into account the corrections to the
asymptotic scaling behavior (\ref{eq:Rscaling}). Including the leading
corrections we have
\begin{equation}
  R_s \approx f(u L^{1/\nu}) + L^{-\omega} g(u L^{1/\nu}),
\label{eq:Rscaling2}
\end{equation}
where $g(x)$ is a scaling function and $\omega$ a universal exponent.
For the standard 3D XY universality class numerical simulations
give~\cite{CHPV-06,PV-02} $\oxy=0.785(20)$.

\subsection{FSS at the transition lines}
\label{nrcb}

\begin{figure}[tb]
  \includegraphics{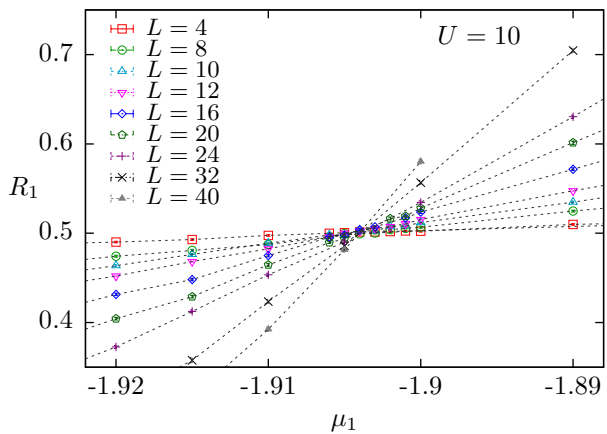}
 \caption{Helicity-modulus combination $R_1$ at $T=1$, $\mu_2=0$,
   $U=10$, as a function of $\mu_1$, close to the normal-to-superfluid
   transition.  }
 \label{U10a}
\end{figure}

\begin{figure}[tb]
 \includegraphics[width=7.5truecm]{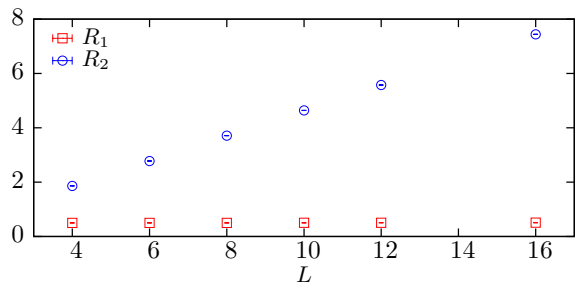}
 \caption{ $R_s=\Upsilon_s L$ for the two gases at the
   normal-to-superfluid transition of gas 1, for $T=1$, $U=4$,
   $\mu_2=0$, and $\mu_1=-1.9035$.  For $L\to\infty$, $R_1$ 
   approaches the XY critical value $R^*=0.516(1)$, while
   $R_2$ increases linearly with $L$ as it is  appropriate
   for a gas in the superfluid phase. 
}
 \label{U10b}
\end{figure}

\begin{figure}[tb]
 \includegraphics{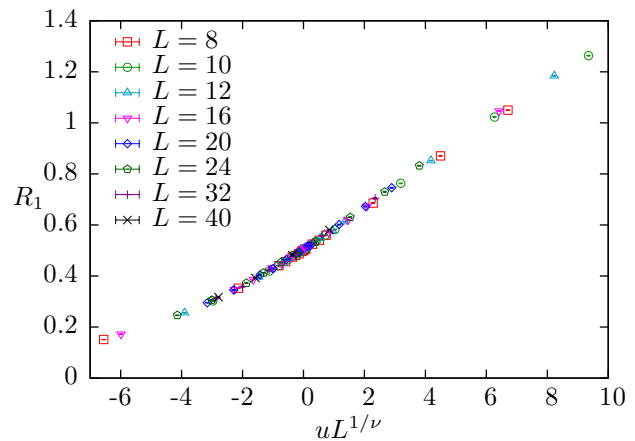}
 \caption{Helicity modulus combination $R_1$ (we plot the 
   same data reported in Fig.~\ref{U10a}) versus
   $uL^{1/\nu}$, $u=\mu_1-\mu_{1c}$, taking $\nu$ equal to the 
   XY value $\nuxy=0.6717$. We use $\mu_{1c}=-1.9035$, obtained by fitting the 
   data to Eq.~(\ref{eq:Rscaling_taylor}).  }
 \label{U10c}
\end{figure}

In the MF approximation, for $\mu_2=0$, $U=10$, and sufficiently small
temperature values, the system undergoes three different phase
transitions as $\mu_1$ is decreased at fixed $T$, see
Fig.~\ref{fig:Tpos_hc}. (i) First, starting from large values of
$\mu_1$, component 1 undergoes a normal-to-superfluid transition,
while component 2 remains in the normal phase. (ii) As $\mu_1$ is
further decreased, also component 2 condenses, while component 1 is in
the superfluid phase.  (iii) Finally, a second normal-to-superfluid
transition of component 1 occurs, but in this case the second
component is superfluid.

Figs.~\ref{U10a}, \ref{U10b} and \ref{U10c} show QMC results at $T=1$
for the transition (iii) at $\mu_1<0$, up to $L=40$.  The estimates of
$R_1$ show a crossing point at $\mu_{1c}\approx -1.904$, see
Fig.~\ref{U10a}, indicating that component 1 undergoes a phase
transition, while $R_2$ increases linearly with $L$, see
Fig.~\ref{U10b}, which is the appropriate behavior expected for a gas
in the superfluid phase.

In order to check that the critical behavior belongs to the 3D XY
universality class, we verify that the data are consistent with
Eq.~(\ref{eq:Rscaling2}), taking the XY values for the critical
exponents. In practice, we fit the data to
\begin{equation}\label{eq:Rscaling_taylor}
R = R^* + \sum_{i=1}^m a_{i} u^i L^{-i/\nuxy} +
                   L^{-\oxy} \sum_{j=0}^n b_{j} u^j L^{-j/\nuxy}, 
\end{equation}
with $u= \mu_1 - \mu_{1c}$. We set $\nuxy=0.6717$ and $\oxy=0.785$,
which are the best available estimates of the two exponents for the 3D
XY universality class.  For our data $u L^{-1/\nuxy}$ is small, so
that we have replaced the scaling functions $f(x)$ and $g(x)$ with
their expansions (to order $m$ and $n$, respectively) around $x=0$.
The values of $m$ and $n$ have been chosen by checking the quality of
the fit and the stability of the results with respect to the order of
the expansions. Around $\mu_1\approx -1.905$, good fits are obtained
by taking $m=1$ or 2 and $n=0$.  Correspondingly, we estimate
$\mu_{1c}=-1.9035(5)$.  The quality of this XY-biased fit can be
checked by plotting $R_1(L;\mu_1)$ versus $u\,L^{1/\nuxy}$ with
$u=\mu_1-\mu_{1c}$, see Fig.~\ref{U10c}. We observe a good collapse of
the data, confirming the XY nature of the transition.

This FSS analysis confirms the conjecture that the critical behaviors
along the normal-to-superfluid transition lines of a single species
belong to the 3D XY universality class, even when the other species is
in the superfluid phase.  Moreover, corrections to scaling always
decay as $L^{-\oxy}$, where $\oxy$ is the leading irrelevant exponent
for the XY universality class. Therefore, the interactions between the
critical and the noncritical component give rise to corrections that
are quite suppressed, decaying at least as fast as $L^{-\oxy}$.

\subsection{The phase diagram for $\mu_2=0$ and $U=4$}
\label{phdia}

\begin{figure}[tb]
 \includegraphics{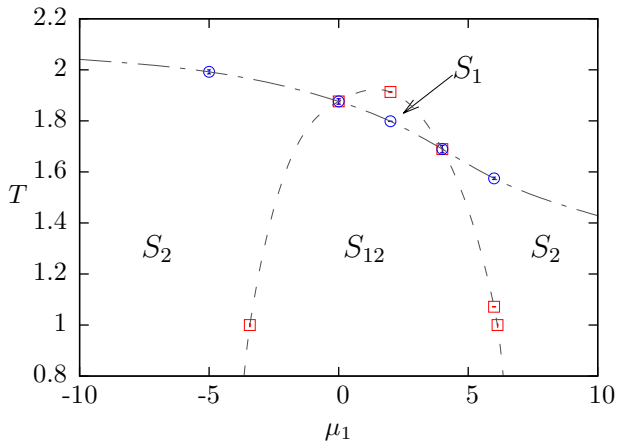}
 \caption{Phase diagram of the 3D 2BH model (\ref{sBH}) in the HC
   limit, for $\mu_2=0$ and $U=4$, as a function of $\mu_1$ and $T$.
   Transitions where component 1 condenses are labelled with squares,
   those where component 2 condenses with circles.  The interpolating
   lines are only meant to guide the eye.  The same phase diagram,
   computed in the MF approximation, is shown in
   Fig.~\ref{fig:Tpos_hc}.  }
 \label{3DphdiaU4}
\end{figure}

\begin{figure}[tb]
 \includegraphics{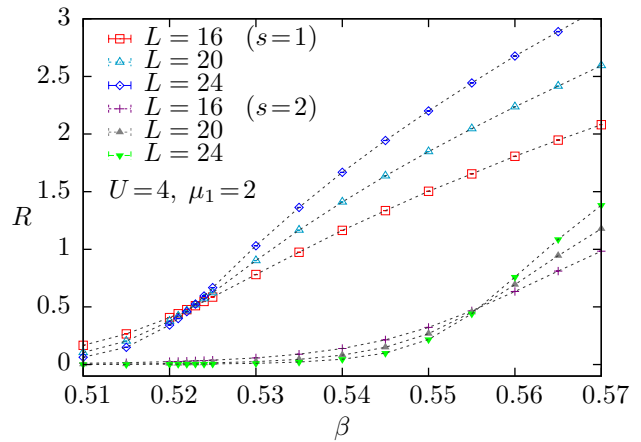}
 \caption{Estimates of $R_1$ and $R_2$ versus $\beta\equiv 1/T$ at
   $\mu_2=0$, $U=4$ and $\mu_1 = 2$. Two different crossing points are
   visible, providing evidence for two distinct transitions.  }
 \label{fig:mua2_decoupling}
\end{figure}

\begin{figure}
    \includegraphics{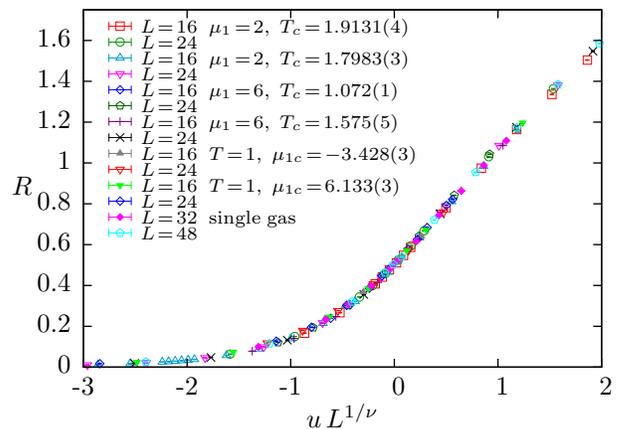}
\caption{FSS plot of the helicity modulus of the condensing species at
  various normal-to-superfluid transitions for $\mu_2=0$ and
  $U=4$. The corresponding critical values in the $T$-$\mu_1$ plane
  are indicated in the labels.  We show data at $\mu_1 = 2$ (at the
  two transitions considered in Fig.~\ref{fig:mua2_decoupling}),
  $\mu_1=6$ (at the normal-to-superfluid transitions of both gases),
  and at $T=1$ at the two transitions of gas 1 driven by $\mu_1$.  For
  comparison, we also report results for the BEC transition of the
  single-species Bose-Hubbard model at $\mu=0$ and in the HC limit.
  The data are plotted versus $uL^{1/\nu}$, where $\nu=\nuxy=0.6717$,
  $u = a(T - T_c)$ for the transitions at fixed $\mu_1$, and $u =
  b(\mu_1- \mu_{1c})$ for those at fixed $T$, see Eq.~(\ref{usca}).
  The constants $a$ or $b$ (they assume different values at each
  transition) are optimized to obtain the best collapse of the data. }
\label{fig:collapse_all_transitions}
\end{figure}

For $\mu_2=0$ and $U=4$ we have repeated the FSS analysis of
Sec.~\ref{nrcb} for other values of the model parameters, with the
purpose of determining an approximate phase diagram, to be compared
with that obtained in the MF approximation, see
Fig.~\ref{fig:Tpos_hc}. Our numerical results are in qualitative
agreement with the MF predictions, confirming the presence of two MCPs
where four transition lines meet, see Fig.~\ref{3DphdiaU4}.

To verify the predicted behavior we have performed simulations at
different values of $\mu_1$, $\mu_1 = -5$, 2, and 6, varying the
temperature $T$.  In Fig.~\ref{fig:mua2_decoupling} we show the
helicity modulus of the two gases at $\mu_1=2$ as a function of $\beta
= 1/T$.  We observe that $R_1$ and $R_2$ cross at two different values
of $T$, indicating the presence of two separate (but close)
normal-to-superfluid transitions. If we move down from the
high-temperature phase decreasing $T$, we first observe the
condensation of gas 1 at $T_{c}=1.9131(4)$ and then that of gas 2 at
$T_{c}=1.7982(3)$. Two different transitions are also observed at
$\mu_1 = 6$. Here, however, the order is reversed. Decreasing the
temperature, first gas 2 condenses at $T_{c} = 1.575(5)$, then gas 1
condenses at $T_c = 1.072(1)$. At $\mu_1 = -5$ we have observed only
one transition, related to component 2. We have also considered a
different line in the phase diagram, keeping the temperature fixed, $T
= 1$, and varying $\mu_1$. In this case we observe two
normal-to-superfluid transitions of the same component, gas 1, at the
boundaries of the superfluid phase $S_{12}$, at $\mu_1=6.133(3)$ and
$\mu_1=-3.428(3)$.

FSS analyses analogous to those described in Sec.~\ref{nrcb} confirm
that all transitions belong to the 3D XY universality class.  As a
further check, in Fig.~\ref{fig:collapse_all_transitions} we show the
helicity modulus close to the transitions we have investigated, as a
function of $u L^{1/\nu}$.  By tuning appropriately the constants $a$
or $b$, cf.  Eq.~(\ref{usca}), at each transition (if our data are
obtained at fixed $\mu_1=\mu_{1c}$ we optimize the constant $a$,
while, if data are obtained at fixed $T=T_c$, we optimize $b$) we
obtain a perfect collapse of the data, confirming the universality of
the scaling function $f(x)$ defined in Eq.~(\ref{eq:Rscaling}).  We
also report the helicity modulus at the BEC transition of a single HC
Bose-Hubbard gas. Results fall on top of those obtained for the
mixture, confirming the XY nature of the transition.

As already anticipated by the MF computations, the phase diagram
reported in Fig.~\ref{3DphdiaU4} has two MCPs, where four transition
lines meet.  At each MCP, both gases simultaneously condense. Their
locations can be inferred from the numerical results of
Ref.~\cite{CNPV-15}, where the finite-temperature BEC transitions of
the 2BH model for two equal bosonic gases were studied.  In
particular, when $\mu_1=\mu_2=0$, the two identical gases condense at
$T_c=1.88(1)$ for $U=4$, and at $T_c=1.69(1)$ for $U=-4$.  Using the
particle-hole relation (\ref{transf}), the latter transition implies
an analogous transition at $\mu_1=4$, $\mu_2=0$, $U=4$, and
$T_c=1.69(1)$. Clearly, the two transitions at $U=4$, $\mu_1=0$,
$\mu_2=0$, $T_{\rm mc}=1.88(1)$ and $U=4$, $\mu_1=4$, $\mu_2=0$,
$T_{\rm mc}=1.69(1)$, must correspond to the MCPs of the phase diagram
reported in Fig.~\ref{3DphdiaU4}, since they are characterized by the
simultaneous BEC of both gases.

As shown in Ref.~\cite{CNPV-15}, the critical behavior of the
transition of two equal HC bosonic gases with on-site inter-species
density-density interaction is controlled by a decoupled XY FP.
However, in this case the competition of the two U(1) order parameters
leads to unusual slowly-decaying scaling corrections. Indeed, the
inter-species density-density interaction gives rise to scaling
corrections that decay very slowly, as $\xi^{\,-0.022}$, where $\xi$
is the diverging length scale at the transition. Such scaling
corrections are not present in standard transitions belonging to the
XY universality class, where they decay as $\xi^{-\oxy}$ with $\oxy
\approx 0.78$.

\section{Multicritical behaviors}
\label{multi}

The competition of distinct types of order gives generally rise to
multicritical phenomena. More specifically, a multicritical point
(MCP) is observed at the intersection of two critical lines
characterized by different order parameters.  Multicritical behaviors
occur in several physical contexts: in anisotropic antiferromagnets,
high-$T_c$ superconductors, multicomponent polymer solutions,
disordered systems, etc., see, e.g.,
Refs.~\cite{LF-72,FN-74,NKF-74,KNF-76,GT-83,LGT-86,
  Zhang-97,Aharony-02,Aharony-02-2,
  CPV-03,HPV-05,CPV-05,PV-05,HS-07,PV-07,HPPV-07}. The scaling
behavior at a MCP is controlled by the stable fixed point (FP) of the
RG flow, which can be studied by field-theoretical approaches based on
the appropriate LGW $\Phi^4$ theory.

\subsection{The U(1)$\oplus$U(1) LGW theory}
\label{lgwth}

\begin{figure}
\vskip5mm 
\includegraphics[width=4.2cm]{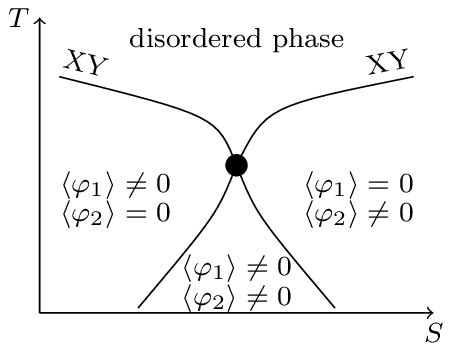}
\includegraphics[width=4.2cm]{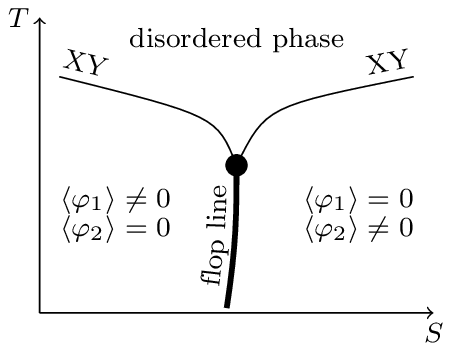}
\caption{Different phase diagrams for models with two different
  interacting U(1) order parameters. In the left panel the MCP is
  observed at the point where four transition lines meet
  (tetracritical point); all transitions are continuous.  In the right
  panel two continuous transitions (thin lines) and one first-order
  transition (thick line) meet at the MCP (bicritical point).  }
\label{mcpd}
\end{figure}

The LGW theory describing the competition of the two different U(1)
order parameters is obtained by constructing the most general $\Phi^4$
theory of two complex fields $\varphi_{s}({\bm x})$, which is
invariant under independent U(1) transformations acting on each of
them.  Its Hamiltonian is
\begin{eqnarray}
{\cal H}_{\rm LGW} &=&\int d^3 x\,\Bigl[
\sum_{s,\mu} |\partial_\mu \varphi_s |^2  +  
\sum_s r_s |\varphi_s|^2 \label{mcphi4s}\\
&+& {1\over 24} \sum_s v_{s} |\varphi_s|^4 
+ {1\over 4} \,u\, |\varphi_1|^2 |\varphi_2|^2\Bigr],
\nonumber
\end{eqnarray}
with two quadratic parameters $r_1$ and $r_2$, and three quartic
parameters $v_{1}$, $v_2$, and $u$.  A multicritical behavior is
obtained by tuning the quadratic parameters $r_1$ and $r_2$
simultaneously to their critical values, keeping the quartic
parameters fixed. Note that the theory is well defined (the quartic
potential is bounded from below) for $v_1 > 0$, $v_2 > 0$, and $u >
-{1\over3} \sqrt{v_1 v_2}$.

Mean-field calculations show that the U(1)$\oplus$U(1) LGW theory
(\ref{mcphi4s}) leads to two different phase
diagrams~\cite{NKF-74,KNF-76,PV-05}, see Fig.~\ref{mcpd}, depending on
the sign of $\Delta\equiv v_{1} v_{2} - 9 u^2$. If $\Delta>0$, four
critical lines meet at the MCP (tetracritical behavior), as in the
left panel of Fig.~\ref{mcpd}, while, if $\Delta<0$, two critical
lines and one first-order line (bicritical behavior) are present, see
the right panel of Fig.~\ref{mcpd}. Note that, in the HC limit, the
2BH model should correspond to the LGW theory with $\Delta>0$, because
of the correspondence $V_s\sim v_s$ and $U \sim u$.  Therefore, we
expect a tetracritical behavior.  A bicritical behavior is expected
instead in the opposite limit $V\lesssim U$.  The MF results presented
in Sec.~\ref{phmf} are completely consistent with this prediction.

\subsection{Multicritical scaling}
\label{mcpsca}

In the LGW theory the transition lines appearing in Fig.~\ref{mcpd}
are obtained by tuning one of the two quadratic parameters $r_1$ and
$r_2$ to its critical value. Multicritical behaviors arise when both
of them are tuned to criticality.  Therefore, generic multicritical
behaviors are associated with two relevant scaling fields $w_1$ and
$w_2$ (analytic functions of the model parameters such that
$w_1=w_2=0$ at the MCP) with positive RG dimensions $y_1$ and $y_2$.
For example, in the case of the 2BH model (\ref{sBH}) $w_1$ and $w_2$
may be taken as linear combinations of the temperature and of the
chemical potentials of the two gases.  In the absence of external
fields, the singular part of the free-energy density is expected to
obey the scaling law
\begin{equation}
F_{\rm sing}(w_1,w_2,w_3,...) = 
b^{-d} {\cal F}(b^{y_1} w_1, b^{y_2} w_2, b^{y_3} w_3,...),
\label{freeenb}
\end{equation} 
where $b$ is an arbitrary {\em blocking} variable.  Here, we have
introduced additional irrelevant scaling fields $w_i$, $i\ge 3$ with
RG dimensions $y_i < 0$, that give rise to scaling corrections at the
critical point.  Neglecting their contribution and appropriately
fixing the arbitrary variable $b$ as $b=|w_1|^{-1/y_1}$, we obtain the
asymptotic scaling expression
\begin{equation}
F_{\rm sing} \approx |w_1|^{d/y_1} f_\pm(w_2 |w_1|^{-y_2/y_1}), 
\label{freeen}
\end{equation} 
where $f_\pm$ are universal scaling functions, which depend on the
sign of $w_1$: $f_+(x)$ should be considered for $w_1 > 0$, $f_-(x)$
in the opposite case.  Close to the MCP, all transition lines
correspond to constant values of the product $w_2 |w_1|^{-y_2/y_1}$.

Within the LGW theory, a standard multicritical behavior can only be
observed if there exists a stable FP for the corresponding RG flow and
the system is in its attraction domain. In the opposite case, the flow
generically runs to infinity and the transition is discontinuous.
Note that this can also occur if a stable FP exists, but the system is
outside its attraction domain.

We should note that the unstable FPs of the theory are also associated
to multicritical behaviors. However, in this case one should perform
additional tunings of the parameters and correspondingly introduce
additional relevant scaling fields.  For example, consider a FP that
is unstable with respect to one of the RG quartic perturbations, i.e.,
such that the flow ends at the FP only if one performs one additional
tuning of the initial parameters.  This means that there is an
additional relevant scaling field.  Eq.~(\ref{freeenb}) still holds,
but now $y_3 > 0$. Therefore, the contribution of $w_3$ can no longer
be neglected approaching the critical point.  From a more
phenomenological point of view, this higher-order multicritical
behavior can be observed by varying three model parameters. In the
corresponding parameter space, one has surfaces of standard critical
transitions. These surfaces intersect along lines that correspond to
standard multicritical behavior. The transition may be continuous,
controlled by the stable FP, or of first-order, if the RG flow goes to
infinity.  The higher-order multicritical behavior is observed at the
intersection of the multicritical lines. Of course, such points are
quite difficult to observe in practice.

\subsection{Perturbative field-theoretical expansion}
\label{MSseries}

The critical behavior at a continuous transition is controlled by the
FPs of the RG flow, which are determined by the common zeroes of the
$\beta$ functions associated with the parameters appearing in the
quartic potential.  The presence of a stable FP controls the universal
features of the critical behavior if the transition is continuous.  If
no stable FP exists, the generic transition is expected to be of first
order.

The $\beta$ functions of the theory can be computed using perturbation
theory. In the calculation one should be careful to tune $r_1$ and
$r_2$ to their critical value to obtain the critical theory. This
requirement is automatically satisfied if one considers the $\epsilon$
expansion, which is based on dimensional regularization around four
dimensions \cite{tHV-72}. Indeed, in this regularization scheme, one
considers directly the massless critical theory. The same is true in
the related 3D scheme of Ref.~\cite{SD-89}, the so-called
$\overline{\rm MS}$ scheme without $\epsilon$ expansion. Here, one
also considers the $\overline{\rm MS}$ perturbative series, but does
not expand in powers of $\epsilon$, setting $\epsilon=1$.

The Hamiltonian fields and parameters are renormalized \cite{ZJ-book}
by setting $\varphi_s = Z_{s}^{1/2} \varphi_{sr}$, $v_{s} = A_d
\mu^\epsilon Z_{v_s}(v_{sr},u_r)$, $u = A_d \mu^\epsilon
Z_{u}(v_{sr},u_r)$, where $v_{sr},\,u_r$ are the $\overline{\rm MS}$
renormalized quartic couplings.  The five renormalization functions
$Z_s$ and $Z_{v_s,u}$ are normalized so that $Z_s\approx 1$, $Z_{v_s}
\approx v_s$ and $Z_u\approx u$ at tree level.  Here $A_d$ is a
$d$-dependent constant given by $A_d\equiv 2^{d-1} \pi^{d/2}
\Gamma(d/2)$.  The $\overline{\rm MS}$ $\beta$ functions are obtained
by differentiating the renormalized couplings with respect to the
scale $\mu$, keeping the bare couplings $v_1$, $v_2$, and $u$ fixed.
The two-loop $\beta$ functions associated with the quartic couplings
are
\begin{eqnarray}
\beta_{v_1} &=& - \epsilon \, v_{1r} + \frac{5}{3} v_{1r}^2 + 3 u_r^2
- \frac{5}{3} v_{1r}^3 - \frac{5}{2} v_{1r} u_r^2 - 6 u_r^3, \nonumber
\\ \beta_{v_2} &=& - \epsilon \, v_{2r} + \frac{5}{3} v_{2r}^2 + 3
u_r^2 - \frac{5}{3} v_{2r}^3 - \frac{5}{2} v_{2r} u_r^2 - 6 u_r^3,
\nonumber \\ \beta_{u} &=& - \epsilon \, u_r + 2 u_r^2 + \frac{2}{3}
v_{1r} u_r + \frac{2}{3} v_{2r} u_r - \frac{5}{2} u_r^3 \nonumber
\\ &&- 2 v_{1r} u_r^2 - 2 v_{2r} u_r^2 - \frac{5}{18} v_{1r}^2 u_r -
\frac{5}{18} v_{2r}^2 u_r .
\label{betat}
\end{eqnarray}
The complete five-loop series are reported in App.~\ref{5loop}.

The zeroes of the $\beta$ functions provide the location of the FPs of
the RG flow. Their stability is controlled by the matrix $\Omega_{i,j}
= \partial \beta_i/\partial g_j$ [the indices correspond to the three
  quartic couplings ${\bf g}\equiv (v_1,v_2,u)$] evaluated at the
given FP. The FP is stable, if all eigenvalues $\omega_i$ of the
stability matrix have positive real part.

\subsection{RG flow and FPs close to four dimensions}
\label{rgfloweps}

We first determine
the FPs and their stability properties close to four dimensions, 
using the first few terms of the standard $\epsilon$ expansion \cite{WF-72}.
We find six different FPs, of which the only stable one is located at
\begin{eqnarray}
&&v_{1r} = v_{2r} = {1\over 2} \epsilon + {7\over 16} \epsilon^2 + O(\epsilon^3),
\nonumber \\
&&u_r = {1\over 6} \epsilon - {1\over 48} \epsilon^2 + O(\epsilon^3).
\label{stableeps}
\end{eqnarray}
In the general O($n_1$)$\oplus$O($n_2$) model this FP is named
biconical FP (BFP) \cite{NKF-74,KNF-76}, it generally satisfies
$v_{1r} \not= v_{2r}$.  In the U(1)$\oplus$U(1) case, however, $v_{1r}
= v_{2r}$ and therefore this FP also appears in the theory with $v_1 =
v_2$ and $r_1 = r_2$,
\begin{eqnarray}
{\cal H}_{\rm LGW} &=& \int d^3x\,\Bigl[
\sum_{s,\mu} |\partial_\mu \varphi_s|^2  +  
r \sum_s |\varphi_s|^2 \label{phi4s}\\
&&+ {v\over 24} \sum_s |\varphi_s|^4 
+ {u\over 4}\, |\varphi_1|^2 |\varphi_2|^2\Bigr],
\nonumber
\end{eqnarray}
which is symmetric under the larger symmetry group ${\mathbb
  Z}_{2,e}\otimes[{\rm U(1)}\oplus{\rm U(1)}]$.  Note that this FP is
degenerate with the O(4) FP at leading order in $\epsilon$.  As a
consequence, $n$-loop calculations at this FP provide results to
$O(\epsilon^{n-1})$ only.  Thus, the available five-loop series
reported in App.~\ref{5loop} allow us to determine the location of the
FP only to order $\epsilon^4$.  For the same reason the smallest
eigenvalue of the stability matrix is of order $\epsilon^2$: $\omega_1
\approx \epsilon^2/6$.

The U(1)$\oplus$U(1) LGW theory reduces itself to the O(4)-symmetric
$\Phi^4$ theory when $r_1=r_2$ and $v_{1}=v_{2}= 3 u$. Correspondingly, the
RG flow has an O(4)-symmetric FP at
\begin{equation}
v_{1r} = v_{2r} = 3 u_r = {1\over 2} \epsilon + {13\over 48} \epsilon^2 + 
O(\epsilon^3)  .
\label{o4eps}
\end{equation}
This FP is unstable in the full theory (\ref{mcphi4s}), since one
eigenvalue of the stability matrix is negative, $\omega_1 \approx -
\epsilon^2/6$. Therefore, it corresponds to a higher-order
multicritical behavior with three relevant scaling fields, of RG
scaling dimensions $y_1 \approx 2 - \epsilon/2$, $y_2 \approx 2 -
\epsilon/6$, and $y_3\approx \epsilon^2/6$.

The LGW theory decouples into two identical U(1) $\Phi^4$ theories
when $u=0$.  We can therefore identify a decoupled FP (DFP) with $u_r = 0$.
At the DFP, the two U(1) order parameters are decoupled, with a 
critical behavior belonging to the XY universality class.
The DFP is located at
\begin{eqnarray}
v_{1r} = v_{2r} = {3\over 5} \epsilon + {9\over 25} \epsilon^2 + 
O(\epsilon^3), \quad u_r = 0.
\label{dfpeps}
\end{eqnarray}
The DFP is stable within each U(1) theory.  Therefore, its stability
properties in model (\ref{mcphi4s}) depend only on the RG dimension
$y_u$ of the coupling $u$ associated with the quartic term
$|\varphi_1|^2|\varphi_2|^2$ that couples the two fields.  The RG
dimension $y_u$ can be evaluated using general scaling
arguments~\cite{Aharony-76,GT-83,Aharony-02,Aharony-02-2}.  At the
DFP, the operator $|\varphi_1|^2|\varphi_2|^2$ scales as the product
of two energy-like operators of the $d$-dimensional XY universality
class.  Therefore, the RG dimension $y_u$ is given by
\begin{equation}
y_u = {2\over \nuxy} - d .
\label{yug}
\end{equation}
Using $\nuxy \approx 1/2
+ \epsilon/10$, we obtain $y_u \approx \epsilon/5 > 0$.
Therefore, the DFP is unstable close to four dimensions.

The other three FPs also have $u_r = 0$. Their stability matrix has
two or three negative eigenvalues, and hence they can only be observed
by tuning four or five different system parameters.  They are of
little relevance for interacting Bose gases.

The above calculations can be straightforwardly extended to
$O(\epsilon^5)$ [$O(\epsilon^4)$ in the case of the stable FP] using
the complete series reported in App.~\ref{5loop}. However, methods
based on the $\epsilon$ expansion allow us to find only those 3D FPs
which can be related, by analytic continuation, to those present close
to four dimensions.  But new FPs may emerge in three dimensions, which
cannot be detected by the $\epsilon$ expansion, because they do not
have a 4D counterpart.  This means that the $\epsilon$ expansion may
not provide the correct description of the 3D RG flow.  For example,
this occurs for the Ginzburg-Landau model, in which a complex scalar
field is coupled to a gauge field, which is appropriate to describe
superconductors and the nematic--smectic-A transition in liquid
crystals \cite{HLM-74}. Although $\epsilon$-expansion calculations do
not find a stable FP \cite{HLM-74}---therefore, they predict a
first-order transition---numerical analyses show that these systems
can also undergo continuous transitions in three dimensions, see,
e.g., Refs.~\cite{MHS-02,NK-03}.  This implies the presence of a
stable FP in the 3D Ginzburg-Landau theory, in agreement with
experiments in liquid crystals~\cite{GN-94}.  Other examples are
provided by the O(2)$\otimes$O($N$) LGW $\Phi^4$ theories describing
frustrated spin models with noncollinear order~\cite{CPPV-04,NO-15},
the $^3$He superfluid transition from the normal to the planar
phase~\cite{DPV-04}, etc...  Therefore, a more conclusive analysis of
the RG flow in three dimensions requires a direct 3D study.

\subsection{Fixed points in three dimensions}
\label{rgflow3D}

We now extend the analysis to the 3D case.  Since the $\epsilon$
expansion suggests that the relevant FPs belong to the plane
$v_{1r}=v_{2r}$, we first consider the 3D FPs that appear in the LGW
theory (\ref{phi4s}), and discuss their stability in the multicritical
theory in which the exchange symmetry is broken.

The analysis of the FPs for model (\ref{phi4s}) is reported in
Ref.~\cite{CNPV-15}.  Two stable FPs are identified: the DFP and a
second FP, named asymmetric FP (AFP).  The DFP controls the
transitions at which two identical gases condense simultaneously. The
AFP, instead, is the relevant FP for transitions at which only one gas
undergoes BEC, breaking the exchange symmetry of model (\ref{phi4s}).

Let us now discuss the stability of these two FP within the
multicritical theory (\ref{mcphi4s}), starting with the DFP.  For $u =
0$ the model corresponds to two noninteracting U(1) systems. We thus
obtain for the RG dimensions of the quadratic operators $y_1 = y_2 =
1/\nuxy = 1.4888(3)$. The two quartic perturbations that are present
for $u =0$ can be identified with the quartic perturbation of the
standard U(1) theory, so that $y_{v_1}= y_{v_2} = -\oxy =
-0.785(20)$. The RG dimension of the perturbation coupling the two XY
models can be computed as in the 4D case, using Eq.~(\ref{yug}) and
$\nuxy=0.6717(1)$ ~\cite{CHPV-06}. We obtain
\begin{equation}
y_u = {2\over \nuxy} - 3 = - 0.0225(4),
\label{yu}
\end{equation}
which is also negative, confirming the stability of the DFP in three
dimensions, at variance with the behavior close to four dimensions.
Note, however, as already discussed in Ref.~\cite{CNPV-15}, that $y_u$
is quite small. Thus, it gives rise to very slowly decaying
corrections, that are quite difficult to detect. Since $u = 0$, we
have $\Delta = v_1 v_2 - 9 u^2 > 0$. Thus, the DFP should be relevant
for systems that have a tetracritical MCP, see Fig.~\ref{mcpd}. The
asymptotic decoupling of the critical modes allows us to simplify
Eq.~(\ref{freeenb}).  Neglecting scaling corrections, i.e. setting
$w_i = 0$ for $i \ge 3$, we can simply rewrite
\begin{eqnarray}
F_{\rm sing}(w_1,w_2) &&\approx b^{-d} {\cal F}_{\textsc{XY}} (b^{y_1} w_1) + 
    b^{-d} {\cal F}_{\textsc{XY}} (b^{y_1} w_2) \nonumber \\
&&\approx 
   a_{\pm}^{(1)} |w_1|^{3\nuxy} + a_{\pm}^{(2)} |w_2|^{3\nuxy},
\end{eqnarray}
where $a_{\pm}^{(a)}$ are constants.  By appropriately choosing $w_1$
and $w_2$, we can additionally set $a_+^{(1)} = a_+^{(2)} = 1$, so
that $a_-^{(1)} = a_-^{(2)}$. The scaling fields are defined by the
requirement that the transition lines correspond to $w_1 = 0$ or $w_2
= 0$.

The second stable FP of the reduced theory (\ref{phi4s}) is the AFP
that controls the critical behavior when only one of two bosonic
species condenses~\cite{CNPV-15}.  It also appears in the
O(2)$\otimes$O(2) LGW theory describing the critical modes of some
frustrated spin models with noncollinear
order~\cite{Kawamura-88,PRV-01,CPPV-04}.  Within the multicritical
theory (\ref{mcphi4s}), the AFP should describe bicritical points
(right panel of Fig.~\ref{mcpd}), essentially because it is associated
with the BEC of only one species.  The RG dimensions of the two
relevant perturbations at the AFP correspond to the dimensions of the
quadratic operators.  The RG dimension $y_1$ is obtained from the
relation $y_1=1/\nu$, where $\nu$ is the correlation-length exponent
of the LGW theory (\ref{phi4s}).  The analysis of the perturbative
expansions within two different renormalization
schemes~\cite{CNPV-15,CPPV-04} (the so-called $d=3$ MZM and
$\overline{\rm MS}$ schemes) gives $\nu=0.57(3)$ and $\nu=0.65(6)$,
respectively, so that $y_1\approx 1.7$. The RG dimension $y_2$ can be
derived from the results reported in Ref.~\cite{CPV-05} for the
quadratic perturbations to the FPs of the O(2)$\otimes$O(2) theory.
We obtain $y_2=1.34(15)$ and $y_2=1.25(4)$ in the two different
perturbative schemes.  The stability properties of the AFP within the
multicritical theory (\ref{mcphi4s}) depend also on the RG dimension
$y_{v_d}$ of the quartic coupling $v_d \sim v_1-v_2$ associated with
the operator $P_d=|\varphi_1|^4 - |\varphi_2|^4$ breaking the exchange
symmetry.  The RG dimension $y_{v_d}$ can be computed by a
perturbative analysis in the $\overline{\rm MS}$ scheme without
$\epsilon$ expansion. We obtain the estimate $y_{v_d} = -0.6(1)$,
where the error takes into account how the estimate changes as the FP
position varies within one error bar (we use the FP estimates of
Ref.~\cite{CPPV-04}) and the dependence on the resummation parameters.
Since $y_{v_d}<0$, the quartic perturbation $P_d$ is irrelevant and
therefore the AFP is stable in the multicritical theory.

Note that the AFP is not connected with the stable biconical FP found
close to 4D.  Indeed, they describe different symmetry-breaking
patterns.  The stable FP close to 4D corresponds to a tetracritical
MCP, in contrast with the AFP, which gives rise to a bicritical
behavior. Apparently, the biconical FP disappears when approaching 3D,
while the AFP, which is absent close to four dimensions, appears only
close to three dimensions \cite{CPPV-04}.

It is also interesting to discuss the O(4) FP, which is already
unstable in the reduced theory (\ref{phi4s}). A complete analysis
shows that there are three different relevant scaling fields at the
O(4) FP in the full theory (\ref{mcphi4s}).  Summarizing, the RG
dimensions of the relevant scaling fields are $y_1=1.333(4)$
associated with the scalar quadratic perturbation at the O(4) FP
(obtained using $\nu=1/y_1=0.750(2)$~\cite{HV-11,Hasenbusch-01}),
$y_2=1.813(6)$ associated with the spin-2 quadratic
perturbation~\cite{CPV-03}, and $y_3=0.125(5)$ associated with the
spin-4 quartic perturbation~\cite{HV-11,CPV-03}.

The analysis we have presented considers only the FPs with $v_{1r} =
v_{2r}$. {\em A priori} other FPs may be present with $v_{1r} \not =
v_{2r}$. As we shall see in the next section, the analysis of the
general RG flow does not find any evidence of additional stable FPs.

\subsection{3D RG flow}
\label{3drgflow}

\begin{figure}
    \includegraphics[angle=-90,width=9truecm]{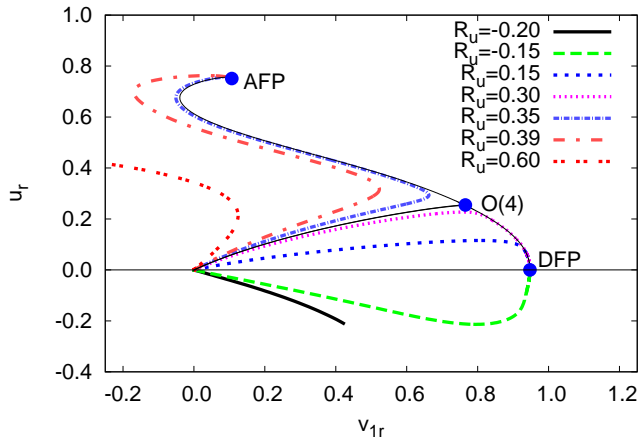}
\caption{RG flow as a function of $v_{1r}$ and $u_r$, for $R_v =1/2$
  and several values of $R_u$, in the $\overline{\rm MS}$ scheme
  without $\epsilon$ expansion.  We also report the position of the
  O(4) unstable fixed point, of the DFP, and of the AFP. Thin
  continuous lines correspond to the separatrices of the RG flow,
  connecting the different FPs.  }
    \label{fig:RGflow-3D}
\end{figure}

In this section we study the RG flow.  In particular, we determine the
RG trajectories starting from the Gaussian FP, where the quartic
couplings vanish. This study allows us to determine the stable FPs and
their attraction domain in the space of the Hamiltonian (bare) quartic
parameters. We use here the $\overline{\rm MS}$ scheme without
$\epsilon$ expansion~\cite{SD-89}, which provides a genuine 3D
critical scheme.

The RG trajectories are obtained by solving the differential equations
\begin{eqnarray}
&&- \lambda {du_r \over d\lambda} = \beta_u[u_r(\lambda),v_{sr}(\lambda)],
\label{betaf}\\
&&  - \lambda {dv_{sr} \over d\lambda} = \beta_{v_s}[u_r(\lambda),v_{sr}(\lambda)],
\nonumber 
\end{eqnarray}
where $s=1,2$,  $\lambda\in [0,\infty)$, with the initial conditions
\begin{eqnarray}
   &&u_r(0) = v_{s\,r}(0) = 0,\\
   && \left. {du_r\over d\lambda}\right|_{\lambda=0} = u, \qquad
   \left. {dv_{sr}\over d\lambda}\right|_{\lambda=0} = v_s.
\nonumber
\end{eqnarray}
Note that the trajectories do not depend on the Hamiltonian parameters
individually, but only through their dimensionless ratios. For this
purpose, we rescale $\lambda \to \lambda/\sqrt{v_1 v_2}$, so that the
initial conditions become $R_u = u/\sqrt{v_1 v_2}$ for
$du_r/d\lambda$, $R_v = \sqrt{v_1/v_2}$ and $1/R_v$ for the
derivatives of ${dv_{1r}/d\lambda}$ and ${dv_{2r}/d\lambda}$,
respectively. Note that the LGW theory (\ref{mcphi4s}) is stable for
$R_u>-1/3$.  One expects tetracritical or bicritical behavior if $R_u
< 1/3$ and $R_u > 1/3$, respectively. Moreover, the symmetry of the
model under $v_1 \to v_2$, $v_2 \to v_1$, allows us to restrict the
analysis to $0 \le R_v \le 1$.  To obtain meaningful results, the
perturbative series are resummed by employing the Pad\'e-Borel
technique, see, e.g., Refs.~\cite{ZJ-book,PV-02}.

Typical results are shown in Fig.~\ref{fig:RGflow-3D}, where we report
a projection of the trajectories in the $v_{1r},u_r$ plane for $R_v =
1/2$ (results for other values of $R_v$ are qualitatively
analogous). For $R_u$ slightly larger than $-1/3$ (see the behavior
for $R_u = -0.20$ in the figure), it is not possible to follow the
flow beyond a certain value of $\lambda$, since the Borel transform
becomes singular on the positive real axis.  These trajectories
clearly correspond to systems that undergo discontinuous
transitions. If we further increase $R_u$, we observe that
trajectories flow to the DFP, which is the stable FP relevant for
small values of $u$.  If $R_u$ is increased again, the relevant FP
changes and the trajectories end up at the AFP. If $R_u$ is further
increased, trajectories run into the non-Borel summable region $v_{1r}
< 0$ and $v_{2r} < 0$ (see the behavior for $R_u = 0.60$ in the
figure).  It is interesting to observe that the range of values of
$R_u$ corresponding to trajectories flowing to the AFP is quite
small. For the approximant shown in the figure, we should have $0.33
\lesssim R_u \lesssim 0.40$. For any $R_u \gtrsim 0.40$ the
trajectories flow to infinity. This suggests that most of the
bicritical MCPs undergo first-order transitions.  Note that the
numerical analysis does not provide any evidence of additional
FPs. Apparently, all relevant FPs belong to the symmetric model with
$v_1 = v_2$.

\section{Conclusions}
\label{conclu}

In this paper we study the critical and multicritical behaviors that
can be observed in 3D mixtures of bosonic gases interacting by
short-range density-density interactions. These systems have a global
U(1)$\oplus$U(1) symmetry, related to independent U(1) transformations
acting on each species.  As a representative of this class of systems,
we consider the 3D Bose-Hubbard model for two lattice bosonic gases
coupled by an on-site inter-species density-density interaction, whose
Hamiltonian is given in Eq.~(\ref{sBH}).  However, the qualitative
features of the finite-temperature phase diagram and the results for
the universality classes associated with the critical and
multicritical behaviors apply to generic bosonic mixtures.

The generic features of the phase diagram of the 2BH model have been
determined in the MF approximation and additionally confirmed by QMC
simulations. The qualititave behavior depends on the model parameters,
such as the chemical potentials and the on-site inter- and
intra-species couplings. By varying them, one can observe several
transition lines, along which one of the two species undergoes a
normal-to-superfluid transition, and different types of multicritical
behavior.

The transition lines separating the different phases generally
correspond to the BEC condensation of one of the two species.  We show
that, independently whether the other species is in the normal or
superfluid phase, the critical behavior of the condensing species
belongs to the 3D XY universality class, characterized by the breaking
of a global U(1) symmetry and short-ranged effective interactions,
which is the same universality class associated with the BEC of a
single bosonic gas. Therefore, the critical modes of the condensing
gas effectively decouple from those of the other species,
independently whether the latter is in the normal or superfluid phase.

The phase diagram of mixtures of bosonic gases also presents
particular points where some transition lines meet. At these points
multicritical behaviors develop, due to the competition of the U(1)
order parameters related to the two bosonic gases.  We investigate
them by a field-theoretical approach based on the effective LGW
$\Phi^4$ theory for two complex scalar fields with global
U(1)$\oplus$U(1) symmetry.  The possible universality classes that
describe the multicritical behaviors are associated with the stable
FPs of the RG flow. They can be determined by studying the RG
trajectories in the critical theory, starting from the unstable
Gaussian FP in the quartic-parameter space.  For this purpose, we
consider the so-called $\overline{\rm MS}$ scheme without $\epsilon$
expansion \cite{SD-89}. We start from the five-loop $\overline{\rm
  MS}$ $\beta$ functions, resum them using the Pad\'e-Borel technique,
and solve the flow equations.  We find two stable FPs, that also
belong to the $\Phi^4$ theory (\ref{phi4s}), which has an additional
$\mathbb{Z}_2$ symmetry related to the exchange of the two order
parameters. This more symmetric model has already been discussed in
the context of the critical behavior of a mixture of two identical
gases \cite{CNPV-15}.  If the system has a tetracritical continuous
transition, see Fig.~\ref{mcpd}, the critical behavior is controlled
by a decoupled FP. Each component shows an XY critical
behavior---correspondingly, the RG dimensions of the two relevant
operators are $y_1=y_2 \approx 1.49$--- but with very slowly-decaying
scaling corrections (they decay as $\xi^{-0.022}$, where $\xi$ is the
correlation length) due to inter-species coupling. If, instead, the
system undergoes a bicritical continuous transition, the critical
behavior is associated with a different asymmetric FP, with $y_1
\approx 1.7$ and $y_2 \approx 1.3$.

Recent experiments on atomic gas mixtures~\cite{PhysRevLett.78.586,
  PhysRevLett.81.1539,Nature.396.345,
  PhysRevLett.82.2228,PhysRevLett.85.2413,PhysRevA.63.051602,
  PhysRevLett.89.053202,PhysRevLett.89.190404,PhysRevLett.99.190402,
  PhysRevLett.99.010403,PhysRevA.77.011603,PhysRevLett.100.210402,
  PhysRevLett.101.040402,PhysRevA.79.021601,
  PhysRevLett.103.245301,PhysRevA.80.023603,PhysRevLett.105.045303,
  PhysRevA.82.033609,PhysRevA.84.011610,NatPhys7.434,NatPhys8.71},
either using two different atomic species or the same atomic species
in two different states, have already obtained several interesting
results on the properties of the low-temperature condensed phase and
on the interplay of the different condensates. They have also
demonstrated the possibility of a robust control of the model
parameters, which may allow the observation of the different phases,
such as those found in the present study, and the determination of the
nature of the critical and multicritical behaviors.  Our results
should provide a complete characterization of the possible BEC
patterns and of the critical behaviors that these systems may develop
along their transition lines.

Most cold-atom experiments have been performed in inhomogeneous
conditions, due to the presence of space-dependent trapping forces,
which effectively confine the atomic gas within a limited space
region~\cite{BDZ-08}. The trapping potential is effectively coupled to
the particle density, which may be taken into account by adding a
further Hamiltonian term to the 2BH Hamiltonian (\ref{sBH}), i.e.,
$H_{\rm trap} = \sum_{s\bm{x}} V_s({\bm x}) n_{s\bm{x}}$ where $V_s$
is the space-dependent potential associated with the external force.
The inhomogeneity arising from the trapping potential introduces an
additional length scale $\ell_t$ into the problem, which drastically
changes the general features of the behavior at the phase transitions.
Experimental data for inhomogeneous trapped cold-atom systems are
usually analyzed using the local-density approximation, see, e.g.,
Ref.~\onlinecite{BDZ-08}.  However, this approach fails to describe
the emergence of large-scale correlations~\cite{ACV-14,CTV-13}. This
problem may be overcome experimentally by using (almost) flat traps,
giving rise to a finite space region where the system is effectively
homogenous~\cite{NGSH-15}.  Otherwise, one may infer the critical
behavior by studying the scaling behavior with respect to the trap
size $\ell_t$, which is expected to be universal and controlled by the
critical exponents of the universality class of the corresponding
homogenous system, in the large trap-size
limit~\cite{CTV-13,CN-14,Pollet-12,CV-09}.

\appendix

\begin{widetext}

\section{Five-loop series of the U(1)$\oplus$U(1) LGW theory}
\label{5loop}

We report here the five-loop perturbative series of the $\beta$
functions used in Sec.~\ref{3drgflow} to analyze the RG flow of the
U(1)$\oplus$U(1) LGW $\Phi^4$ theory.  We consider the perturbative
expansions obtained in $4-\epsilon$ dimensions, using the dimensional
regularization and in the modified minimal-subtraction ($\overline{\rm
  MS}$) scheme.  They were computed in Ref.~\cite{CPV-03} for general
O($n_1$)$\oplus$O($n_2$) theories, but they have never been reported.
Apart from the first few orders, coefficients are reported with a
$10^{-6}$ numerical precision, although they are computed in terms of
fractions and $\zeta$ functions (the exact series are available on
request).  To simplify the formulas, the renormalized couplings are
named $v_1$, $v_2$ and $u$ instead of $v_{1r}$, $v_{2r}$, and $u_r$.
The five-loop $\beta$ functions read
\begin{eqnarray}
&&\beta_{v_1}(v_1,v_2,u) = - \epsilon v_1 + \frac{5}{3} v_1^2 + 3 u^2 -
\frac{5}{3} v_1^3 - \frac{5}{2} v_1 u^2 - 6 u^3  \label{beta1}\\
&&\quad +14.381 u^4 + 36.1747 v_1 u^3 + 8  v_2 u^3+2.3125 v_1^2
u^2+0.125 v_2^2 u^2-1.08333 v_1 v_2 u^2+4.99347 v_1^4 \nonumber \\
&&\quad -120.062 u^5-164.419 v_1 u^4-91.2808 v_2 u^4-165.154 v_1^2
u^3-10.9914 v_2^2 u^3-18.2281 v_1 v_2 u^3-5.32929 v_1^3 u^2 \nonumber \\
&&\quad -0.489598 v_2^3 u^2 
-0.144409 v_1 v_2^2 u^2
+0.969685 v_1^2 v_2 u^2-21.9072 v_1^5 
+ 1090.1 u^6+1633.45 v_1 u^5+947.957 v_2 u^5
\nonumber\\
&&\quad +1140.29 v_1^2
u^4+257.256 v_2^2 u^4 
+608.06 v_1 v_2 u^4+950.994
v_1^3 u^3+37.2454 v_2^3 u^3+32.3079 v_1 v_2^2
u^3+35.291 v_1^2 v_2 u^3 \nonumber\\
&&\quad
+6.68047 v_1^4 u^2+1.27762
v_2^4 u^2-0.142137 v_1 v_2^3 u^2+1.6941 v_1^2
v_2^2 u^2+2.20787 v_1^3 v_2 u^2+120.141 v_1^6\;,
\nonumber\\
&&\beta_{v_2}(v_1,v_2,u) = \beta_{v_1}(v_2,v_1,u) \;,
\label{beta2}\\
&& \beta_{u}(v_1,v_2,u) = - \epsilon u + 2 u^2 + \frac{2}{3} v_1 u +  \frac{2}{3} v_2 u
- \frac{5}{2} u^3 - 2 v_1 u^2 - 2 v_2 u^2  - \frac{5}{18} v_1^2 u - \frac{5}{18} v_2^2 u 
\label{beta3}\\&&\quad
+11.7312 u^4+11.3082 v_1 u^3+11.3082 v_2 u^3+3.50134 v_1^2
u^2+3.50134 v_2^2 u^2+1.11111 v_1 v_2 u^2+0.652778
v_1^3 u \nonumber\\&&\quad
+0.652778 v_2^3 u
 -85.8801 u^5-87.0641 v_1 u^4-87.0641 v_2 u^4-38.5894 v_1^2
u^3-38.5894 v_2^2 u^3-32.9072 v_1 v_2 u^3
\nonumber\\&&\quad
-12.3167
v_1^3 u^2-12.3167 v_2^3 u^2-2.06356 v_1 v_2^2
u^2-2.06356 v_1^2 v_2 u^2-1.99192 v_1^4 u-1.99192 v_2^4 u
+ 711.585 u^6\nonumber\\&&\quad
+896.552 v_1 u^5
+896.552 v_2 u^5+455.484 v_1^2
u^4+455.484 v_2^2 u^4+507.235 v_1 v_2 u^4
+176.847
v_1^3 u^3+176.847 v_2^3 u^3
\nonumber\\&&\quad
+95.0588 v_1 v_2^2
u^3+95.0588 v_1^2 v_2 u^3+54.8793 v_1^4 u^2+54.8793
v_2^4 u^2
+7.87236 v_1 v_2^3 u^2+4.76209 v_1^2
v_2^2 u^2
\nonumber\\&&\quad
+7.87236 v_1^3 v_2 u^2+7.99517 v_1^5
u+7.99517 v_2^5 u\;.
\nonumber
\end{eqnarray}

\end{widetext}

\begin{acknowledgements}         
We acknowledge  
computing time at   the Scientific Computing Center of
INFN-Pisa.  We acknowledge funding from  FP7/ERC starting grant No. 306897.
\end{acknowledgements}


\begin{thebibliography}{99}
\bibitem{footnote-meanfield} The MF Hamiltonian
  \eqref{eq:mf_hamiltonian} is in general an infinite-dimensional
  matrix so that, to obtain numerical results, one must fix a cutoff
  on the site occupancy.  We fix both occupation numbers of the two
  species to be at most $N_\mathrm{max}$, so that
  \eqref{eq:mf_hamiltonian} becomes a $(N_\mathrm{max}+1)^2 \times
  (N_\mathrm{max}+1)^2$ real symmetric matrix.  Of course, the cut-off
  is arbitrary and may induce a bias in MF calculations.  For this
  reason, the stability of the results must be checked a posteriori,
  repeating the computation for increasing values of $N_\mathrm{max}$.
  Note that, in the HC limit, no truncation is needed and the MF
  hamiltonian is simply a $4 \times 4$ matrix.
\end{thebibliography}
\end{document}